# Respiratory Motion Correction in Abdominal MRI using a Densely Connected U-Net with GAN-guided Training

Wenhao Jiang[1], Zhiyu Liu[1], Kit-Hang Lee[1], Shihui Chen[2], Yui-Lun Ng[1], Qi Dou[3], Hing-Chiu Chang[2], and Ka-Wai Kwok[1]

[1]Department of Mechanical Engineering, The University of Hong Kong, Hong Kong
`kwokkw@hku.hk`
[2]Department of Diagnostic Radiology, The University of Hong Kong, Hong Kong
[3]Department of Computing, Imperial College London, London, UK

**Abstract.** Abdominal magnetic resonance imaging (MRI) provides a straightforward way of characterizing tissue and locating lesions of patients as in standard diagnosis. However, abdominal MRI often suffers from respiratory motion artifacts, which leads to *blurring* and *ghosting* that significantly deteriorates the imaging quality. Conventional methods to reduce or eliminate these motion artifacts include breath holding, patient sedation, respiratory gating, and image post-processing, but these strategies inevitably involve extra scanning time and patient discomfort. In this paper, we propose a novel deep-learning-based model to recover MR images from respiratory motion artifacts. The proposed model comprises a densely connected U-net with generative adversarial network (GAN)-guided training and a perceptual loss function. We validate the model using a diverse collection of MRI data that are adversely affected by both synthetic and authentic respiration artifacts. Effective outcomes of motion removal are demonstrated. Our experimental results show the great potential of utilizing deep-learning-based methods in respiratory motion correction for abdominal MRI.

## 1    Introduction

Clinical magnetic resonance imaging (MRI) often suffers from blurring and ghosting, namely motion artifacts, which are mainly caused by physiological motion of patients. During the magnetic resonance (MR) image reconstruction [1], such motion gives rise to *k*-space mismatches due to any variable phase errors taken place in phase-encoding steps. The resultant artifacts would seriously reduce the quality of images and further affect diagnostic accuracy. Motion prevention, artifact reduction, and motion correction are common approaches to reduce/eliminate these artifacts. The first one is simply by instructing the patients to hold their breath and keep their body still throughout the scanning. This requires cooperation with the patient, but inevitably causes patient discomfort. It would be further problematic in pediatric imaging which is why general anesthesia [2] is often employed. However, its use carries the inherent risks associated with general anesthesia, and also requires the presence of anesthetists, increasing total cost.

A straightforward approach to artifact reduction is by shortening the imaging time so that it is much lower than the duration of body motion. Techniques such as echo planar imaging (EPI) [3] and compressed sensing (CS-MRI) [4] are widely applied to improve the immunity against motion. However, there would be drawbacks such as geometric distortion or limited spatial resolution of the acquired images. Besides, these methods even introduced new types of artifacts [5]. Triggering and gating [6] are also widely adopted to reduce periodic artifacts caused by physiological motion from heartbeats and breathing. The scan time is able to be selected within breathing positions or cardiac cycles when the corresponding motions are detected to be minimal. However, such motion cycles may be irregular in clinical reality, particularly for those patients with arrhythmia and free-breathing scenarios. The removal of artifacts is still very challenging.

Post-processing or retrospective correction could alleviate the aforementioned problems. With the imaging data acquired, the correction process requires integration of navigator data, as well as implementation of large number of iterative algorithms [7]. Therefore, the required exhaustive computation is the primary barrier of using such post-processing correction. It may be resolved by the recent boost of calculation capability. There has been a trend in recent studies to utilize the highly parallel computing power of graphic processing units (GPU) by introducing deep-learning-based methods.

Tamada *et al*. [8] proposed a multi-channel convolutional neural network (CNN) based method to reduce motion artifacts caused by respiration. In this work, respiration-induced artifacts were simulated by adding phase errors in the phase encoding direction. Simulated images were trained and tested via an 8-layer CNN. Armanious *et al*. [9] compared the rigid and non-rigid motion correction results of four GAN-based models based on T1-weighted spin echo (SE) sequences from 17 subjects. Nevertheless, comparisons of these learning-based methods on diverse sequences of MRI scans with both synthetic and authentic motion artifacts have not yet been studied.

In this work, we propose a dense U-Net based model that incorporates the GAN-guided training strategy and perceptual loss to correct respiratory motion of abdominal MRI. Promising motion correction results with both synthetic and authentic respiratory artifacts are demonstrated in the testing phase. The major work contributions are: **i)** The dense U-Net based model is pre-trained with realistically simulated motion on the open access TCGA-LIHC dataset [10], and then fine-tuned on experimental MR images affected by human respiration; **ii)** Comparisons among the proposed and other deep-learning-based methods on abdominal MRI motion correction; **iii)** Validation on diverse experimental datasets (GRE, FSE) along with the real motion pairs.

## 2 Methods

### 2.1 Network Structure

The network structure for motion correction is proposed as shown in **Fig. 1**. We designed a GAN-based model as data synthesizer to take advantage of its generalization capability on small datasets. The model comprises three major parts: **i)** A Dense U-Net [11] generator, **ii)** A Deep CNN model discriminator, and **iii)** A pre-trained VGG19 feature extraction network. The generator is used to produce clean images from motion affected images, concurrently to preserve perceptual details for diagnostic purposes. The discriminator is trained to distinguish generated motion-free image from real clean images. The perceptual network is implemented to extract high-level features from generated images and real images, which were used to optimize the perceptual loss.

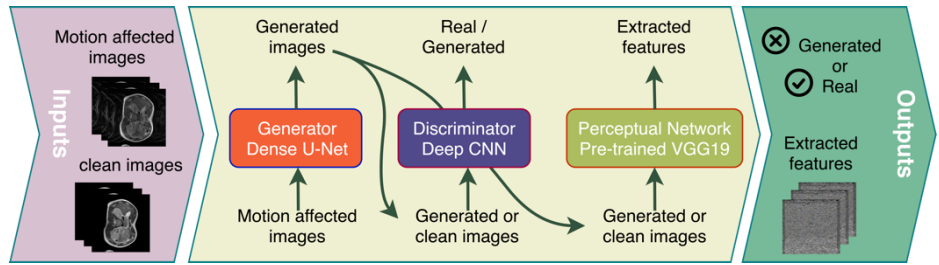

**Fig. 1.** Schematic of the proposed motion correction framework.

**Fig. 2** illustrates the architecture of our Dense U-Net generator. The network consists of four densely connected blocks connected by three transition layers and five upsampling layers. Such densely connected blocks prevent the model from acquiring repetitious features, and also from maximizing information flow for searching the optimal solution. The transition layers act as a compression unit to avoid feature maps from over-expanding. Note that bilinear interpolation is employed in the upsampling layers.

The discriminator comprises *eight* convolution layers, *two* dense layers and a final sigmoid activation. The convolution layers are ordered by a 1-stride and a 2-stride in turn. The first *two* layers have *sixty-four* 3×3 filter kernels and the number of filter kernels is doubled for every *two* subsequent layers. Loss functions such as pixel-wise mean square error (MSE) often result in loss of details while maintaining global structures. This is because high frequency contents were usually neglected when achieving high PSNR. Inspired by the SRGAN [12], we apply a perceptual loss function capable of measuring the MSE of features that are extracted by an ImageNet pretrained VGG-19 network. In this way, differences between high-level features of the generated and real images could be minimized so as to refine the anatomical structures of images.

## 2.2 Perceptual Loss Function

Instead of using pixel-wise MSE which led to *blobby* and overly smooth results (to be discussed in session **3.4**), we incorporate a perceptual loss function in this model. The perceptual loss is a weighted sum of a content loss and an adversarial loss:

$$L_p = L_c + 10^{-3} L_a \tag{1}$$

where $L_p$ stands for *perceptual* loss, $L_c$ and $L_a$ represent *content* and *adversarial* loss, respectively. The content loss is defined by the Euclidean distance between feature representation of real motion-free image and generated clean image from motion affected images. The loss function is defined as:

$$L_c = \frac{1}{H \times W} \sum_{x=1}^{W} \left\{ \sum_{y=1}^{H} \left[ \phi(I_c)_{xy} - \phi\left(G_{\theta_G}(I_m)\right)_{xy} \right]^2 \right\} \tag{2}$$

where the $W$ and $H$ are, respectively, the width and height of VGG19 feature maps. The real clean and motion-affected images are $I_c$ and $I_m$, respectively. The generated motion-free image is denoted as $G_{\theta_G}(I_m)$, while the feature representation $\phi$ is the activated convolution inside the VGG19 network. The adversarial loss is applied to direct the generator to output natural images that is indistinguishable by the discriminator, defined as below:

$$L_a = -\sum \log \left\{ D_{\theta_D} \left[ G_{\theta_G}(I_m) \right] \right\} \tag{3}$$

where $D_{\theta_D}\left[G_{\theta_G}(I_m)\right]$ is the probability of natural image referring to discriminator.

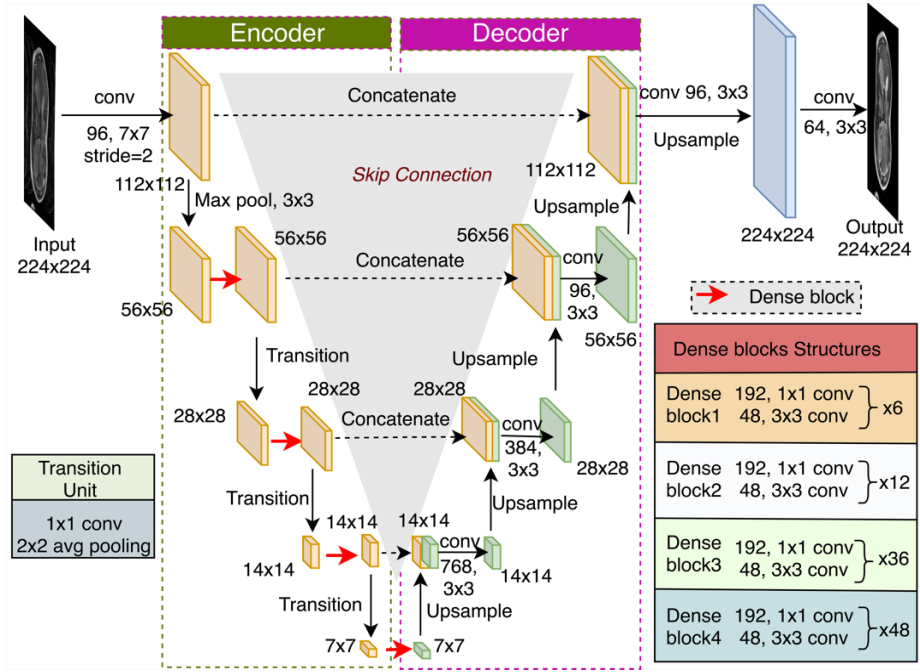

**Fig. 2.** Schematic of the dense U-Net generator network. Decoders were skip-connected to encoders. In each dense block, outputs of each convolutional layers were densely connected with all the successive layers.

## 3 Experiments

### 3.1 Data Acquisition and Respiration Simulation

The motion-free and motion-corrupted image pairs have to be precisely matched. To acquire such experimental dataset, it inevitably prolonged the scanning time, as each subject has to be scanned twice at least. Two datasets were acquired from *eight* healthy subjects on 1.5T MRI scanner (GE HDxt). Two MR sequences, gradient echo (GRE) and a fast spin echo (FSE), were employed to acquire abdomen images. The scanning parameters are set as: TR = 50 ms, matrix size = 128×128 mm for GRE and 192×192 mm for FSE, FOV = 189×189 mm for GRE and 284×284 mm for FSE, number of slices = 14 (GRE) and 20 (FSE), slice thickness = 8 mm. In GRE sequence, each subject went through *two* scans, under breath-free and breath-hold conditions. In FSE sequence, the paired motion-free and motion-corrupted images were acquired, respectively, using respiratory gating and under breath-hold condition. Each subject was scanned for approximately 60 min on average, which is varied by the different respiration conditions of individual volunteer. To augment the training data, the magnitude images from each coil were extracted from paired motion-free and motion-corrupted dataset. The training dataset consists of 1,164 paired images from *six* subjects. 406 paired images from *two* subjects were used for validation. To improve the model generalization capability, we conducted the training on the open access dataset, TCGA-LIHC [10], with simulated respiration. The dataset contains 1,688 MRI, PT and CT series from 97 patients. We applied the simulation method referring to [8] in order to generate the *k*-space datasets with respiration-induced phase error in the phase-encoding direction.

**Fig. 3.** Comparison of our proposed model with other *three* well-known deep learning models in motion correction. Residual motion images were calculated by subtracting the target patches from the motion-affected or model output pictures. Darker residual motion images indicate better correction results.

### 3.2 Implementation Details

The model was implemented in python with Keras [13] high-level neural networks API. We applied an Adam [14] optimizer with the learning rate of $2\times10^{-4}$ and a batch size of 5. Grayscale MR images were converted into RGB to match the pre-trained VGG-19 network. To augment the training data and alleviate the overfitting problem, we randomly applied mirror flip on pictures with the probability of 0.5. The training of the proposed model required approximately 18 hours on an Nvidia RTX 2080Ti GPU with 12GB memory. In the test phase, the inference time of one image ranges from 25 to 40 ms on the same GPU.

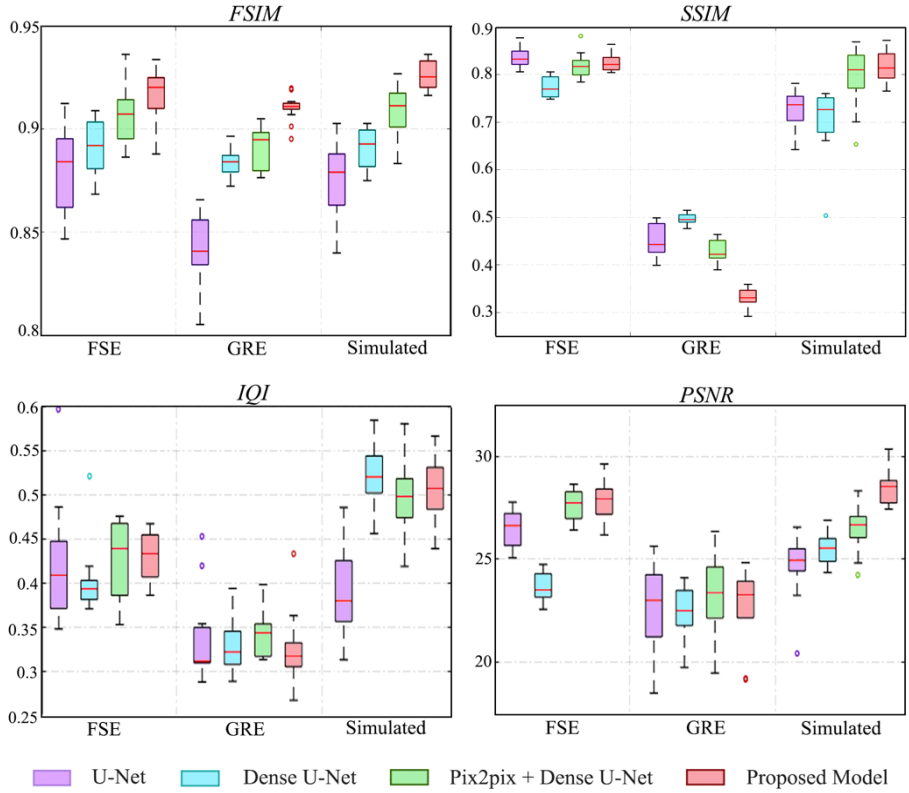

**Fig. 4.** Quantitative comparisons among our proposed model and other deep learning models, regarding to FSIM (*Upper left*), SSIM (*Upper right*), IQI (*Lower left*) and PSNR (*Lower right*). Such artifacts were induced under the FSE, GRE sequences, and the scans with simulated motion.

### 3.3 Experimental Results

The results of the proposed model and other deep-learning-based methods are compared qualitatively (in **Fig. 3**) and quantitatively (in **Fig. 4**). We applied the conventional U-Net [15], Dense U-Net [11] and pix2pix [16] GAN with Dense U-Net as generator for comparisons. All of these models were trained and tested on the same dataset.

Several observations can be made from **Fig. 3**. For the U-Net model, the model is only capable of removing bulk ghosting outside of the central area of images. The Dense U-Net surpasses it in capturing anatomical contours. However, the organ details are still *blobby* and *blurry*. This problem was alleviated by introducing a pix2pix-GAN-guided training schema, but the textures were yet to be depicted, still leaving some striations in the residual motion images. The proposed model outperformed the others by restoring the sharpest details from motion-affected input images.

The comparisons among the proposed model and the other deep learning models with boxplot figures is demonstrated in **Fig. 4**. Four criteria: 1) Feature Similarity Index (FSIM), 2) Structural Similarity Index (SSIM), 3) Image Quality Index (IQI), and 4) Peak Signal-to-Noise Ratio (PSNR) were selected for our comparison. Tests were conducted on synthetic and authentic respiration artifacts. The proposed model performs the best average FSIM values of 0.920, 0.910 and 0.928 on FSE, GRE and simulated motion, respectively. The relatively short box plots of our proposed model prove the consistency of its outputs. The model also shows competitive and stable performances in SSIM, IQI and PSNR. However, all networks performed poorly on GRE scans. We believe images with low SNR may contribute to the worse performance achieved by our proposed model. Without the 180° pulse, GRE cannot refocus the phase accumulation due to the static-field inhomogeneities, thus causing the acquired images with low SNR.

## 4     Conclusion and Discussion

This paper proposed a novel model incorporating the widespread densely connected U-Net with a GAN-guided training schema and perceptual loss function. Our work is the *first* to apply this model in abdominal MRI motion correction. We were also the *first* to train and test the proposed model on MR images corrupted by both synthetic and authentic respiratory motion. The experimental results on FSE, GRE and simulated-motion datasets achieve an FSIM of 0.920, 0.910 and 0.928 respectively, significantly outperform other deep-learning-based models.

In the future work, we intend to employ 4D MRI models that visualizes precise anatomical structure in continuous time. We will manipulate with the adjacent *k*-space slices in time axis of the 4D images and synthesize more realistic motion artifacts for model training. By doing so, we expect to enhance the generalization capability of the model by introducing more training data. Another further study is to recover signal voids. Although bulk *ghosting* and *blur* could be efficiently removed, signal cancellation is still hard to restore and will leave void areas on the output images (as discussed in Session **3.4**). We will explore the restorative capability of GAN-based-networks by processing on *k*-space or MRI raw data.

**Acknowledgments.**
This work is supported by the Research Grants Council (RGC) of Hong Kong (Ref. No.: 17202317, 17227616, 17206818, 27209515), and Innovation and Technology Commission (Ref. No.: UIM/353).

## References

1. Zaitsev, M., Maclaren, J., Herbst, M.: Motion artifacts in MRI: a complex problem with many partial solutions. Journal of Magnetic Resonance Imaging 42(4), 887-901 (2015).
2. Malviya, S., Voepel-Lewis, T., Eldevik, O. P., Rockwell, D. T., Wong, J. H., Tait, A. R.: Sedation and general anaesthesia in children undergoing MRI and CT: adverse events and outcomes. British journal of anaesthesia 84(6), 743-748 (2000).
3. Mansfield, P.: Multi-planar image formation using NMR spin echoes. Journal of Physics C: Solid State Physics 10(3), L55 (1977).
4. Lustig, M., David D., John M. P.: Sparse MRI: The application of compressed sensing for rapid MR imaging. Magnetic Resonance in Medicine: An Official Journal of the International Society for Magnetic Resonance in Medicine 58(6), 1182-1195 (2007).
5. Zhang, T., Chowdhury, S., Lustig, M., et al.: Clinical performance of contrast enhanced abdominal pediatric MRI with fast combined parallel imaging compressed sensing reconstruction. Journal of Magnetic Resonance Imaging 40(1), 13-25 (2014).
6. Ehman, R. L., McNamara, M. T., Pallack, M., Hricak, H., Higgins, C. B.: Magnetic resonance imaging with respiratory gating: techniques and advantages. American journal of Roentgenology 143(6), 1175-1182 (1984).
7. Loktyushin, A., Nickisch, H., Pohmann, R., Schölkopf, B.: Blind retrospective motion correction of MR images. Magnetic resonance in medicine 70(6), 1608-1618 (2013).
8. Tamada, D., Kromrey, M. L., Onishi, H., Motosugi, U.: Method for motion artifact reduction using a convolutional neural network for dynamic contrast enhanced MRI of the liver. arXiv preprint arXiv:1807.06956 (2018).
9. Armanious, K., Küstner, T., Nikolaou, K., Gatidis, S., Yang, B.: Retrospective correction of Rigid and Non-Rigid MR motion artifacts using GANs. arXiv preprint arXiv:1809.06276 (2018).
10. Erickson, B. J., Kirk, S., Lee, Y., et al.: Radiology Data from The Cancer Genome Atlas Liver Hepatocellular Carcinoma [TCGA-LIHC] collection. The Cancer Imaging Archive (2016). http://doi.org/10.7937/K9/TCIA.2016.IMMQW8UQ
11. Li, X., Chen, H., Qi, X., Dou, Q., Fu, C.W., Heng, P.A.: H-denseunet: Hybrid densely connected unet for liver and tumor segmentation from ct volumes. IEEE transactions on medical imaging 37(12), 2663-2674 (2018).
12. Ledig, C., Theis, L., Huszár, F., et al.: Photo-realistic single image super-resolution using a generative adversarial network. In: Proceedings of the IEEE conference on computer vision and pattern recognition, pp. 4681-4690 (2017).
13. Keras, https://github.com/fchollet/keras, last accessed 2019/04/02.
14. Kingma, Diederik P., Jimmy Ba.: Adam: A method for stochastic optimization. arXiv preprint arXiv:1412.6980 (2014).
15. Ronneberger, O., Fischer, P., Brox, T., et al.: U-net: Convolutional networks for biomedical image segmentation. In: Navab N., Hornegger J., Wells W., Frangi A. (eds.) MICCAI 2015. LNCS, vol. 9351, pp. 234-241. Springer, Cham (2015). https://doi.org/10.1007/978-3-319-24574-4_28
16. Isola, P., Zhu, J.Y., Zhou, T., Efros, A.A.: Image-to-image translation with conditional adversarial networks. In: Proceedings of the IEEE conference on computer vision and pattern recognition, pp. 1125-1134 (2017).